\documentstyle[floats,twocolumn,prl,aps]{revtex}
\begin{document}
\draft
\def\ref{\par\noindent\hangindent=3mm\hangafter=1}
\narrowtext
%########################################################################
%                    TITRE ET AUTEURS                                   #
%########################################################################
{
\title{
Density of States Driven Anisotropies
Induced by Momentum Decoupling
in High-$T_c$ Superconductors}
}
\author{G. Varelogiannis, A. Perali, E. Cappelluti and L. Pietronero}
\address{
Dipartimento di Fisica,
Universita di Roma "La Sapienza", Piazzale Aldo Moro 2, I-00185 Roma,
Italy}
%########################################################################
%                          ABSTRACT.                                    #
%########################################################################
\author{\parbox{397pt}{\vglue 0.3cm \small
Momentum decoupling arises when small momentum transfer
processes dominate
the electron-phonon scattering and implies that anisotropies
in superconductivity
are driven by the anisotropies of the density of states.
Considering an
isotropic s-wave interaction in
the momentum decoupling regime we give a natural simultaneous
explanation to various aspects of
ARPES and tunnel experiments on $Bi_2Sr_2CaCu_2O_8$, including
the correlation of gap magnitude and visibility of the
dip above the gap,
the enhancement of anisotropy with temperature,
the presence of gap minima away from the $\Gamma-X$ direction and
a gap maximum in the $\Gamma-X$ direction,
the similarity of tunnel and ARPES spectra in the
$\Gamma-\bar{M}$ direction and the asymmetry in
the SIN tunnel spectra where the dip structure is present only at
negative sample bias. }}
\maketitle
\par

The observation by Angle Resolved Photoemission Spectroscopy (ARPES)
of the superconducting gap
in $Bi_2Sr_2CaCu_2O_8$
\cite{Hwu,Shen,Capunzano,Science,Reports} failed in answering
the controversial question of the symmetry of the order
parameter.
Some experimental results support the d-wave hypothesis
\cite{Shen}, while others point to a rather complex mixed state
in which the gap has nodes away from the $\Gamma-X$ direction
\cite{Capunzano} or even has no nodes at all \cite{Hwu,Science}.
These contradictions, together with the
absence of gap observation in $YBa_2Cu_3O_7$ and the weakness of the
gap values reported on $Bi_2Sr_2CaCu_2O_8$ \cite{Reports},
illustrate the difficulty in measuring
the gap with the ARPES technique.

However
there are
some qualitative points on which all ARPES experiments
are in agreement. The first point is that the higher values of the
gap (which are the easier to measure) are reached in the direction
where the density of electronic states (DOS) on the Fermi level  is maximal,
and in general the gap value is smaller when the DOS is smaller.
The second important point is that the dip structure above the gap
is more visible in the direction where the gap and the DOS are
maximal, and that the visibility of the dip structure follows the variations
of the gap and DOS. A third important remark is that from the ARPES
experiments in the optimal direction for the gap,
one obtains a spectral function
very similar to the tunnel spectra
\cite{tunnel}. In the case of tunneling we see the average of the
spectral function over the Fermi surface, and the similarity of
ARPES and tunnel data in so anisotropic materials is quite
surprising.
In addition,
some very important
experimental trends have been reported in recent ARPES and tunnel experiments.
In Ref.
\cite{Science}
it is shown that
the anisotropy is strongly enhanced when we
move from the $T=0$ regime to the $T\rightarrow T_c$ regime.
In Ref. \cite{Capunzano} it has been shown that the
gap has minima about $10^o$ away from the $\Gamma-X$ direction
and a maximum in the $\Gamma-X$ direction.
On the other hand,
detailed vacuum tunneling spectroscopy measurements \cite{Renner}
report an asymmetric density of states in $Bi_2Sr_2CaCu_2O_{8+\delta}$,
with the dip
%bove the gap
appearing only at negative sample bias.

In the following, we will give a simultaneous explanation to
all the previously cited experimental points.
We will see in particular that all the previous points
characterize a superconductor in which the
isotropic s-wave electron-phonon (or other boson) coupling is
dominated by forward scattering processes.
Dominance of forward scattering processes
can be due to the vicinity of a phase separation instability
\cite{GriCa}, that appears reasonable in
models describing strongly correlated electronic systems
\cite{Crete}.
On the other hand, there is evidence from Raman scattering
\cite{Cardonna} that small momentum transfer (Raman active) phonons are
strongly affected by superconductivity in cuprates.
Notice that a somewhat analogous situation
has already been considered
by Abrikosov in order to understand the change
of sign of the order parameter reported in some experiments on
$YBa_2Cu_3O_7$ \cite{Abrikosov}, and
that dominance of forward scattering is an optimal condition
for a positive contribution of non-adiabatic effects to superconductivity
\cite{LP1,LP}.

When forward scattering is dominant, there is ``Momentum Decoupling'' (MD)
in the
superconducting behavior, implying a different coupling in different
regions of the Fermi surface.
Indeed one has to keep in mind that the coupling is an {\it intensive}
quantity
and not an extensive one.
In the case of MD
the coupling at each region of the Fermi
surface is proportional to the local DOS, and therefore
{\it
the anisotropies in the superconducting state
are induced by the anisotropies of the density of states in the
normal state}.

Let us now illustrate briefly how the MD
appears for small momentum transfer
processes and why it is the {\it only} situation which
leads to DOS dependent anisotropies.
The anisotropic Eliashberg equation in the off-diagonal sector, for
an Einstein spectrum
can be written with usual notations as follows.
%$
%\Delta(\vec{k},i\omega_n)Z(\vec{k},i\omega_n)= \pi T
%\sum_{m}
%\int_{S_F}{d^2 p \over S_F}
%N(E_F,p)
%$
%$
%{|g(\vec{k}-\vec{p})|^2\Omega\over
%\Omega^2+(\omega_n-\omega_m)^2}
%{ \Delta(\vec{p},i\omega_m)\over
%\sqrt{ \omega_m^2+\Delta^2(\vec{p},i\omega_m)}}
%\eqno(1)
%$
$$
\Delta_{\vec{k}}Z_{\vec{k}}=
{\pi T\over S_F}\sum_{m}
\int_{S_F}
{d^2 pN(E_F,\vec{p})|g(\vec{k}-\vec{p})|^2\Omega\over
\Omega^2+(\omega_n-\omega_m)^2}
F(\Delta_{\vec{p}},\omega_n)
\eqno(1)
$$
The $\vec{k}$ dependence is contained in the coupling
$|g(\vec{k}-\vec{p})|^2$.

In conventional s-wave superconductors it is
assumed that the interaction $|g(\vec{k}-\vec{p})|^2$
is almost
constant on the Fermi surface and it leads to
an isotropic
On the other hand,
if one supposes that $|g(\vec{k}-\vec{p})|^2$ has a relevant momentum
dependence in the vicinity of the Fermi surface (as it is the case
in the d-wave scenario where this function reflects electron-spin
fluctuation coupling) then we obtain a $\vec{k}$-dependent coupling
and from equation (1) one can obtain an anisotropic gap.
However
the anisotropy of the superconducting parameters
is mainly imposed by the anisotropy
of the interaction and {\it not} from
the anisotropy of the density of states.
In order to obtain significant DOS induced anisotropies
one has to consider an
{\it isotropic s-wave interaction}
dominated by forward scattering processes.
This can be illustrated by taking for example an interaction
which is sharply peaked
at zero momentum $|g(\vec{k}-\vec{p})|^2\approx
g^2\delta(\vec{k}-\vec{p})$. Then from equation (1), it is easy to
see that there is {\it momentum decoupling}.
We obtain a momentum independent Eliashberg equation
which provides the gap function $\Delta(\vec{k},i\omega_n)$
for each momentum $\vec{k}$ on the Fermi surface.
This last equation is analogous to the
isotropic Eliashberg equation {\it with a coupling strength
proportional to the density of states at the given point
of the Fermi surface} $N(E_F,\vec{k})$.

Of course a $\delta$-function peak at $q=0$ is a rather unrealistic
coupling function.
However MD occurs even
for finite $q$ provided $q$ {\it is small compared to the characteristic
momentum of the DOS variations over the Brillouin zone}.
To illustrate this point we performed numerical simulations on a simple
two-dimensional
BCS model \cite{later}. In that case the gap is given by
$$
\Delta(\vec{k})=\sum_{\vec{p},|\xi_p|<\Omega_D}
%(\vec{k}-\vec{p})
{-V(\vec{k}-\vec{p})
\Delta(\vec{p})\over 2\sqrt{\xi^2_{\vec{p}}+\Delta^2(\vec{p})}}
\tanh\biggl(\sqrt{{\xi^2_{\vec{p}}+\Delta^2(\vec{p})\over 2T}}\biggr)
\eqno(2)
$$
We consider an isotropic s-wave electron-phonon
coupling having at small momenta a Lorentzian
behavior as a function of the norm of the exchanged momentum
$
V(\vec{q})=-V\bigl(1+|\vec{q}|^2/q_c^2\bigr)^{-1}
$
In this spectrum the electron-phonon scattering is dominated
by the processes which transfer a momentum smaller than $q_c$.

For clarity, we will consider here \cite{later}
the
simple nearest neighbor tight binding dispersion
at half-filling $\xi_{\vec{k}}=-2t[cos(k_x)+cos(k_y)]$
(the lattice spacing is taken equal to unity).
The Fermi surface is
a square defined by
$k_x=k_y\pm \pi$ and $k_x=-k_y\pm \pi$ with saddle points
at $(0,\pm \pi)$ and $(\pm \pi,0)$. The minimum of the density of
states is obtained at the points $(\pm \pi/2,\pm\pi/2)$ and
therefore the characteristic length of the DOS variations over the
Brillouin zone is $\pi/\sqrt{2}$. We expect therefore that for
$q_c>\pi/\sqrt{2}$ the gap might be isotropic while for
$q_c$ sufficiently smaller than $\pi/\sqrt{2}$  MD
should manifest leading
to DOS induced anisotropies.
In fact
in figure 1 we show
the ratio of the gap at
$(0,\pi)$ over the gap at the points where the
DOS is minimal $(\pi/2,\pi/2)$ as a function of $q_c$.
We can see that for $q_c < \pi/\sqrt{2} $ this ratio begins to
be appreciably different from unity indicating
the onset of a DOS induced anisotropy because of MD.
%t is clear that for finite transfered momenta it is
%ossible to obtain significant DOS induced anisotropies.

We are now going to see how MD can explain
all the features of the ARPES and tunnel
experiments mentioned in the introduction.
We take first the temperature dependence
of the anisotropy reported in \cite{Science}.
%e shoed previously that MD
%eans an independent Eliashberg equation for each $\vec{k}$ point on
%he Fermi surface. In Ref. \cite{Science} is reported that
%he gap in the $\Gamma-X$ direction goes to zero at a temperature
%maller than $T_c$.
%The system behaves like if superconductivity at
%\Gamma-\bar{M}$ is not influenced by superconductivity at
%\Gamma-X$
%hich is the precise implication of MD.
If MD was perfect in $Bi_2Sr_2CaCu_2O_{8+\delta}$,
the temperature at which the gap disappears
in the $\Gamma-X$ direction
should be smaller to that in
the $\Gamma-\bar{M}$ direction.
In fact since the DOS is smaller in this direction,
the coupling and $T_c$ should also be smaller.
Therefore, in the case of MD the anisotropy is
enhanced close to $T_c$, and if
MD would be perfect it should even diverge.

We will now show how within our BCS model the results of
\cite{Science} can be
qualitatively reproduced when finite (small) momenta are transfered.
In figure 2 we report the temperature dependence of
$\Delta(0,\pi)$ and $\Delta(\pi,\pi)$ for different values
of $q_c$, and
in figure 3 we give the corresponding temperature dependence of the
anisotropy ratio $R=\Delta(0,\pi)/\Delta(\pi,\pi)$.
The critical temperatures are obtained by solving
numerically the hermitian
eigenvalue problem of the linearized equations near $T_c$ \cite{later}.
%or each one of the points, several hours of CPU time in a standard
%nix workstation are necessary \cite{later}. The regularity of
%ur points reflects the quality of the numerical convergence.

When
$q_c=\pi/4$ the DOS induced
anisotropy, because of partial MD, is already significant ($R\approx 1.7$)
but the anisotropy is almost temperature independent (Fig. 3).
For
smaller values of $q_c$ there is a continuous deformation
of the $T$ dependence of $\Delta(\pi,\pi)$ \cite{later}
from the large $q_c$ regime
to the $q_c\rightarrow 0$ regime
where, as expected in perfect MD, $\Delta(\pi,\pi)$ should have
a BCS behavior going to zero at a temperature of the order $T_c/2$,
and therefore the anisotropy ratio should diverge close to $T_c$.
The results of Ref. \cite{Science} point to a strong MD
regime. But one should bear in mind that, if the small gap is smaller
than the temperature at which it is measured, it becomes experimentally
inaccessible \cite{critical}. Taking into account these damping
effects neglected in our BCS model, the results of
Ref. \cite{Science} can be qualitatively understood even
with $q_c$ of the order
$\pi/ 10$ \cite{later}. {\it The enhancement of anisotropy
with temperature is an evidence of
%sotropic
%-wave interaction in the
MD},
%egime},
and \underline {cannot} be understood
in the case of anisotropic interactions like
those considered in the d-wave scenario.

It is difficult to conclude on the value of the coupling from
the gap measurements made by ARPES, because of their large uncertainties.
However there is a qualitative feature
common to all experiments that
certifies that indeed moving from the $\Gamma-\bar{M}$ to the
$\Gamma-X$ direction on the Fermi surface
we go from a strong coupling regime to
a weak coupling regime. In fact in the $\Gamma-\bar{M}$
direction one can observe
a dip structure above the gap that
is a strong coupling effect
independent on the spectral structure of the boson \cite{RC}.
Such a dip appears when for sufficiently strong couplings the
gap $\Delta$ is comparable to the boson energies that mediate
superconductivity ($2\Delta/T_c\geq 5.5$ ).
The stronger is the coupling, the sharper and deeper
is the dip \cite{RC}.
It is a common trend of all experiments
that, moving from $\Gamma-X$ to $\Gamma-\bar{M}$, the visibility of the
dip follows the enhancement of the gap and DOS, indicating the
presence of different couplings at different regions of the Fermi surface
in agreement
with the MD picture.

In Ref. \cite{Capunzano}, it has been reported that the gap
has minima around $30^o-35^o$
and $60^o-65^o$ (angles measured from $X,Y-\bar{M}$)
and a smaller maximum at $45^o$.
The authors claim that there must be a difference between the
results
in the $\Gamma X$ and $\Gamma Y$ quadrants (which is not apparent from
their data) and that the minima
might correspond to nodes. To our analysis instead, these results
will be shown to be an
{\it evidence that gap anisotropies are indeed driven by DOS
anisotropies, indicating that the interaction is isotropic s-wave in the
MD regime}.

We show in figure 4a, the angular dependence
of the density of states
(measuring angles exactly as in Ref. \cite{Capunzano})
for the simplest next nearest neighbor tight binding
dispersion that accounts qualitatively for the $CuO$
bands seen by ARPES \cite{later}.
Fixing the distance from the bottom of the
band at $\approx 350 meV$ as in the experiment \cite{Reports}, we
consider three different characteristic situations
depending on the
distance of the van-Hove singularity from the Fermi level $\delta E$
($\delta E=10meV$ , $40meV$ and $90meV$).
We can see that the DOS has minima at around
$30^o$ and $60^o $ and a maximum at $45^o$ {\it just due to the
buckling of the Fermi surface}.
To obtain the corresponding anisotropies of the gap, we performed
strong coupling calculations for an Einstein phonon
spectrum assuming perfect MD.
We choose a coupling factor that reproduces in the
$\Gamma-\bar{M}$ direction the experimentally reported dip structure
($\lambda\approx 3$) \cite{later}.
The absolute value of the gap depends on the considered
phonon frequency, that is why we show in figure 4b only the relative
variations of the gap. Considering however $\Omega\approx 40 meV$ as
the study of the gap ratio spectral dependence \cite{PRB}
and Raman experiments \cite{Cardonna} indicate, we
reproduce the gap and $T_c$ of these materials \cite{PRB}.

The local minima of the gap at $\approx 30^o$ and $60^o$ \cite{Capunzano}
reflect the local minima of the DOS.
%%%%%%
%%%%%%
%%%%%%
{\it Only in the case of
an isotropic s-wave interaction in the MD regime,
fine structures of the DOS anisotropy like the local
maximum at $45^o$, are reflected in the
gap anisotropy}. If the interaction were not isotropic
(as for example in the case of d-waves), the
anisotropies of the interaction would completely
dominate the fine structures of the DOS anisotropies and the local
maximum at $45 ^o$ should be absent.
As for the magnitude of the anisotropy, this is very dependent on the
distance of the Van Hove singularity to the Fermi level \cite{later}.
In fact the experimental uncertainty on the
value of $ \delta E$ \cite{Reports} is such that one can easily accommodate
the gap anisotropies reported by both Refs. \cite{Science} and
\cite{Capunzano}, and the momentum dependence of the dip visibility
\cite{later}.

Finally, with the concept of MD
we can also understand the qualitative similarity of
the tunnel data and the ARPES data in the direction optimal for
superconductivity. In the case of MD, the tunnel spectrum is
a sum of {\it independent} contributions from various parts
of the Fermi surface, and it does {\it not} reflect an averaged
superconducting behavior \cite{later}.
The tunnel spectra are dominated by the
contribution of the optimal part around $\Gamma-\bar{M}$, since
the
Van-Hove singularity below $E_F$ is extended and covers
about $ 30 \% $ of the Brillouin zone
and the coupling is much stronger in this region.
With this picture we can naturally understand the asymmetry
of the tunnel spectra of Ref. \cite{Renner}.
In fact, the dip structure is seen only at negative sample bias,
because the Van Hove singularity at $\Gamma-\bar{M}$
is {\it below} the Fermi level. Measuring at positive sample bias,
the dynamic behavior reflects the density of states
above the Fermi level (as in inverse Photoemission).
The presence of the dip at negative sample bias and its
absence at positive sample bias \cite{Renner}, indicates that
the density of states at an energy of the order of $\Delta$
%the optimal gap in the $\Gamma-\bar{M}$ direction)
above the Fermi level, is at least $30\%$ smaller to that
at an energy $\Delta$ below the Fermi level,
and this can be easily obtained given the presence of the
Van Hove singularity in the $\Gamma-\bar{M}$ direction.
Because the DOS is smaller above the Fermi surface, the
coupling is smaller and the dip is no more visible \cite{RC,later}.

We acknowledge interesting discussions with
M. Onellion,
Ch. Renner, \O. Fischer, Z.-X. Shen, G. Margaritondo, J. Ma,
M. Campuzano, P.B. Allen,
C.Castellani, C. Di Castro and M. Grilli.

%newpage

%newpage

%\Large \bf Figure Captions}
%vskip 1.5cm

{\bf Figure 1:} Evolution of the anisotropy ratio as a function
of the characteristic range of the
exchanged momenta $q_c$. For $q_c<\pi/\sqrt{2}$ it increases
sharply indicating the onset of MD.
%vskip 1.0cm

{\bf Figure 2:} Temperature dependence of the gap $\Delta$ in the
$(\pi,0)$ (circles) and in the $ (\pi,\pi)$ (triangles) directions
for three characteristic
ranges of exchanged momenta $q_c$.
%vskip 1.0cm

{\bf Figure 3:} Temperature dependence of the anisotropy ratio
for $q_c=\pi/4$ (circles), $q_c=\pi/12$ (triangles) and
$q_c=\pi/20$ (squares). The increase with temperature of this
ratio is a clear indication of MD.
%vskip 1.0cm

{\bf Figure 4:} (a): $N(E_F)$ as a function of the angle $\phi$
measured from the
$X-\bar{M}$ direction for a simple second nearest neighbors
tight binding model with a Van Hove singularity at
$10 meV$ (full line),  $40 meV$ (dashed line) and $90 meV$ (dotted line)
below the Fermi level. (b): The corresponding
gap anisotropy when the Van Hove singularity is at $10 meV$ (triangles),
$40meV$ (circles) and $90 meV$ (squares) below the Fermi level,
obtained by strong coupling calculations assuming
perfect MD.

\end{document}